# The morphology and temperature dependent tensile properties of diamond nanothreads

Haifei Zhan[1], Gang Zhang[2,*], John M. Bell[1], and Yuantong Gu[1,**]

[1]School of Chemistry, Physics and Mechanical Engineering, Queensland University of Technology (QUT), Brisbane QLD 4001, Australia
[2]Institute of High Performance Computing, Agency for Science, Technology and Research, 1 Fusionopolis Way, Singapore 138632

**Abstract:** The ultrathin one-dimensional $sp^3$ diamond nanothreads (NTHs), as successfully synthesised recently, have greatly augmented the interests from the carbon community. In principle, there can exist different stable NTH structures. In this work, we studied the mechanical behaviours of three representative NTHs using molecular dynamics simulations. It is found that the mechanical properties of NTH can vary significantly due to morphology differences, which are believed to originate from the different stress distributions determined by its structure. Further studies have shown that the temperature has a significant impact on the mechanical properties of the NTH. Specifically, the failure strength/strain decreases with increasing temperature, and the effective Young's modulus appears independent of temperature. The remarkable reduction of the failure strength/strain is believed to be resulted from the increased bond re-arrangement process and free lateral vibration at high temperatures. In addition, the NTH is found to have a relatively high bending rigidity, and behaves more like flexible elastic rod. This study highlights the importance of structure-property relation and provides a fundamental understanding of the tensile behaviours of different NTHs, which should shed light on the design and also application of the NTH-based nanostructures as strain sensors and mechanical connectors.

*Corresponding author. Email: zhangg@ihpc.a-star.edu.sg; yuantong.gu@qut.edu.au

## 1. Introduction

Past decades have witnessed huge interests from both scientific and engineering communities on the carbon-based nanostructures, such as fullerenes (0D), carbon nanotube (CNT, 1D) [1], diamond nanowire (DNW, 1D) [2], and graphene (2D) [3]. Their intriguing chemical and physical properties have enabled them as versatile and excellent integral parts for the next generation of devices [4, 5] or multifunctional materials [6] (from 1D nano-fibers/yarns [7, 8] to 2D nanomesh [9], and 3D porous structures [10]). Particularly, driven by their high elastic modulus [11], strength-to-weight ratio, chemical inertness, high thermal conductivity, and relatively easy functionalization, the $sp^3$ bonded DNWs have received an increasing research focus [12-14]. Studies have shown that DNWS have appealing applications as energy absorbing material under UV laser irradiation [15], high efficiency single-photon emitters (with stable and room-temperature operation) [16], and DNA sensing [17, 18].

The attractive usages of DNW have motived researchers to seek effective ways to fabricate/synthesis DNWs with different sizes [19]. Very recently, a new 1D $sp^3$ diamond nanostructure has been reported, termed as diamond nanothread (NTH) [20], which is synthesized through the slow decompression of crystalline benzene in large volume high-pressure cells. Essentially, the diamond NTH is a close-packed $sp^3$-bonded tubular carbon structure, which can be regarded as hydrogenated (3,0) CNTs connected with Stone-Wales (SW) transformation defects [21]. The SW transformation defects interrupt the tubular structure of the diamond NTH. Generally, the diamond NTH is similar to the ultra-thin DNW as formed inside the CNT from diamantine dicarboxylic acid [22].

In fact, different 1D thread-like $sp^3$ C-H polymers have been proposed previously from different perspectives, e.g., tube (3,0) [21], polymer I [23], and polytwistance [24, 25]. Encouraged by this experimental success, several systematic theoretical studies have been carried out to predict other possible atomic structures of NTHs, By enumerating the hexavalent bonding geometries of the benzene molecules, Xu *et al* [26] have identified 50 topologically distinct NTHs, 15 of which are within 80 meV/carbon atom of the most stable member. Excellent mechanical property has been observed in one of the possible diamond NTHs via first-principles calculation[19]. However, considering such diversity in the NTH family, it is of great interest to know how the mechanical properties of the NTH will differ from each other. Furthermore, the effect of temperature and dynamical information about the bond-rearrangement process under strain still remain elusive. In this work, we explore the mechanical properties of NTHs through large-scale molecular dynamics simulations. We find that the mechanical properties of NTH can vary vastly from each other, and the temperature has a significant influence on their mechanical performance.

## 2. Methods

Various NTHs were constructed by varying the bonding patterns between benzenes, which leads to the formation of pentagon, hexagon, heptagon and octagon carbon rings. The 15 most stable NTH members can be classified into three groups including achiral (six models), stiff-chiral (four models) and soft-chiral (five models).[26] In this work, we selected one representative NTH from each group, denoted as NTH-I (achiral), NTH-II (stiff-chiral) and NTH-III (soft-chiral), respectively. The atomic configurations of these three NTHs are illustrated in Figure 1. It is seen that these NTHs have totally different morphologies, i.e., NTH-I shows a zigzag structure, NTH-II appears more like a tube, while NTH-III has a helical morphology. The mechanical properties of different diamond NTHs were acquired through a series of tensile tests performed using large-scale molecular dynamics (MD) simulations.

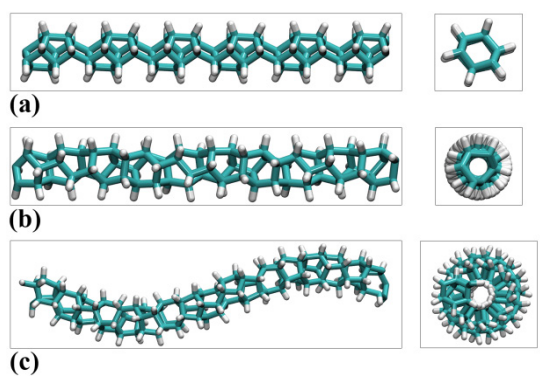

**Figure 1** Atomic configurations of the representative NTHs: (a) achiral NTH-I; (b) stiff-chiral NTH-II; (c) soft-chiral NTH-III. Left shows the view perpendicular to the axis and right shows the cross-sectional view.

For comparison purposes, a similar initial length was chosen for each sample (with periodic boundary conditions in the length direction), i.e., 23.9 nm for NTH-I, 23.5 nm for NTH-II, and 22.4 nm for NTH-III. To initiate the simulation, the widely used adaptive intermolecular reactive empirical bond order (AIREBO) potential was employed to describe the C-C and C-H atomic interactions [27, 28]. This potential has been shown to represent well the binding energy and elastic properties of carbon materials. The C-C cut-off distance was chosen as 2.0 Å. The structures were firstly optimized by the conjugate gradient minimization method and then equilibrated using Nosé-Hoover thermostat [29, 30] for 2 ns (under isothermal-isobaric ensemble). To limit the influence from thermal fluctuations, a low temperature of 1 K was adopted initially. The tensile deformation was achieved by applying a low constant strain rate (namely, $10^{-7}$ fs$^{-1}$) to the fully relaxed structure, and the structure is relaxed for 5 ps after every 0.1% strain increment in the sample. The simulation was continued until the failure of the NTH. A small time step of 0.5 fs was used for all above calculations with all MD simulations being performed using the software package *LAMMPS* [31].

During the tensile simulation, the commonly used virial stress was calculated, which is defined as [32]

$$\Pi^{\alpha\beta} = \frac{1}{\Omega}\left\{-\sum_i m_i v_i^\alpha v_i^\beta + \frac{1}{2}\sum_i \sum_{j \neq i} F_{ij}^\alpha r_{ij}^\beta\right\} \quad (1)$$

Here, $\Omega$ is the volume of the system; $m_i$ and $v_i$ are the mass and velocity of atom $i$; $F_{ij}$ and $r_{ij}$ are the force and distance between atoms $i$ and $j$; and the indices $\alpha$ and $\beta$ represent the Cartesian components. Considering the large morphology difference among the studied samples, we adopted the linear atom density ($\lambda$, in the unit of atoms/Å) to calculate the volume of the structure following Xu *et al* [26], which is about 2.41, 2.45 and 2.91 atoms/Å for the three NTHs, respectively. Such approach has been previously utilized to characterize the (3,0) and (2,2) sp$^3$ carbon tubes [21], and is further systematized for the elastic moduli calculation of nanoscale materials [33]. With the linear atom density, the cross-sectional area of the structure can be approximated by $\lambda V_0$. Here $V_0$ is a reference atomic volume for carbon atom in bulk diamond, which is about 5.536 Å$^3$/atom [21]. To note that adopting different approaches to calculate the cross-sectional area of the NTH would yield to different absolute values of stress, while it will not influence the scaling behaviours as we focused in this paper. According to Eq. 1, the engineering stress is calculated after the relaxation process after each 0.1% strain increment. Correspondingly, we derived the engineering strain based on the applied constant strain rate.

## 3. Results and discussions

### *3.1 Structural influence*

Initially, we assess the tensile behaviours of these NTHs. As compared in Figure 2, the NTHs possess different stress-strain curves, whereas, all NTHs exhibit brittle behaviour, i.e., the stress experiences a sudden drop after continuously increasing to a threshold value. According to the atomic configurations, the NTH starts to fail after passing the threshold value. Thus, this threshold stress is regarded as the failure strength and the corresponding strain is designated as the failure strain. From Figure 2, NTH-I and NTH-III have similar failure strain, i.e., ~ 16% and ~ 18%, respectively. In comparison, a much larger failure strain is observed from NTH-II, which is about 22%. Similarly, NTH-II shows the highest failure strength (around 141 GPa), followed by NTH-I (~ 86 GPa). As expected, the soft-chiral type NT-III exhibits the lowest failure strength of about 79 GPa. The effective Young's modulus, which is a placeholder for the tensile stiffness, is also extracted from the stress-strain curve using linear regression. Based on the assumption of linear elasticity, the initial linear regime with the strain up to 3% has been selected for the fitting, as is widely applied to evaluate the mechanical properties of nanomaterials in previous studies [19, 34-36]. Consistent with the failure strength/strain, NTH-II shows the highest Young's modulus (about 1.09 TPa), followed by NTH-I (about 0.66 TPa) and NTH-III (~ 0.29 TPa). Of interest, we compare the estimated Young's modulus with other one-dimensional carbon allotropies, i.e., CNT and the monoatomic carbyne. According to Liu *et al* [37], the monoatomic carbyne chain, has a Young's modulus around 1.3 TPa, which is close to the NTH-II, but much higher than NTH-I and NTH-III. Similarly, the single-wall CNT is reported to possess a Young's modulus around 1 TPa [38], close to our estimate for NTH-II.

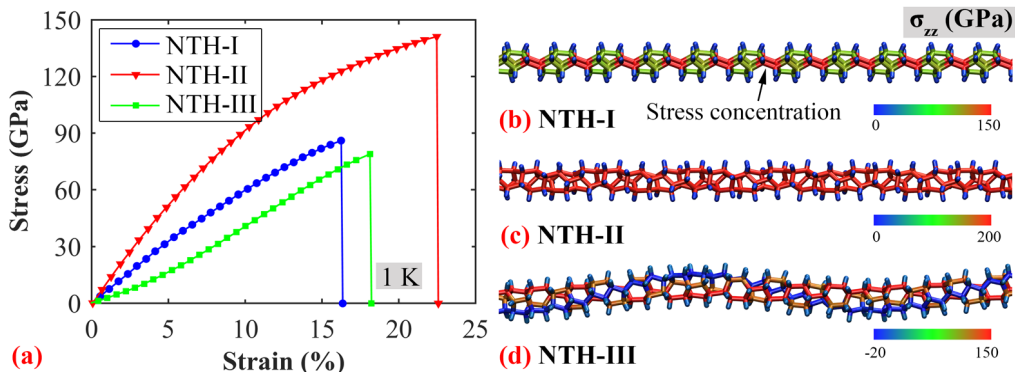

**Figure 2** (a) Comparisons of the stress-strain curves of the three NTHs at 1 K; (b-d) The atomic stress distribution of the three NTHs at the strain of 12% (before failure) along the stretch direction. Atoms are coloured according to the atomic stress along the length direction.

It is of great importance to correlate the mechanical properties of the NTH with its structure. As revealed in Figures 2b, 2c and 2d, the three NTHs exhibit totally different stress status due to their different morphologies. Specifically, the zigzag shape NTH-I shows a stress concentration at the connecting carbon bonds between the two coupled pentagonal carbon rings (Figure 2b). Such stress concentration regions uniformly occur along the length direction, and the failure of the NTH is initiated from these regions with increasing strain. In comparison, the stiff-chiral NTH-II shows a generally even stress distribution pattern during tensile deformation (Figure 2c). This observation is reasonable as NTH-II is actually a polytwistane, which has a uniform structure with its carbon skeleton analogous to (2,1) carbon nanotube [39]. The most striking feature is that the helical NTH-III shows a double-helix stress distribution pattern as plotted in Figure 2d. Specifically, one of the two carbon helixes is experiencing a strong tensile stress (i.e., absorbing most of the tensile strain), with the other one under minor compressive stress state. With increasing strain, failure is triggered along the helix with tensile stress.

It is worthy to mention that the loading rate might influence the tensile behaviours of the diamond NTH. In this regard, we examined the tensile behaviour of the NTH under different strain rates (ranging from $5\times10^{-8}$ to $4\times10^{-7}$ $fs^{-1}$, NTH-III is taken as the representative sample). As illustrated in Figure 3a, the stress-strain curves almost overlap with each other. In the meanwhile, a same stress pattern is found in all examined strain rates (Figure 3b). These results signify that the strain rate ($10^{-7}$ $fs^{-1}$) considered in this work exerts ignorable impacts on the tensile behaviour of the studied NTH, and it is suitable for the investigation purpose. Particularly, although NTH-III is helical in nature, it does not appear to "uncoil" during the stretch, which is also not seen by using other tensile loading schemes, i.e., imposing a constant velocity ($5\times10^{-6}$ Å/fs) to one end of the NTH with another end being fixed (see Supporting information). Such observation is consistent with the relatively linear stress-strain curves from the onset of stretching (Figures 2a and 3a). In all, both NTH-I and NTH-III show stress concentrations during tensile deformation, whereas, NTH-II exhibits a uniform stress distribution. Such observation explains the above finding that Young's modulus of NTH-II is remarkably higher than its counterparts (NTH-I and NTH-III). Considering their tailorable structures, these results suggest a highly tunable tensile mechanical property of NTH. For instance, the structure of the examined NTH-I is like a chain purely made from Stone-Wales transformation defects, which can be changed by introducing hydrogenated (3,0) tubes [19]. Many opportunities are expected for constructing NTHs with designed tensile properties through either altering the structural "defects" or constructing heterojunctions between different NTHs.

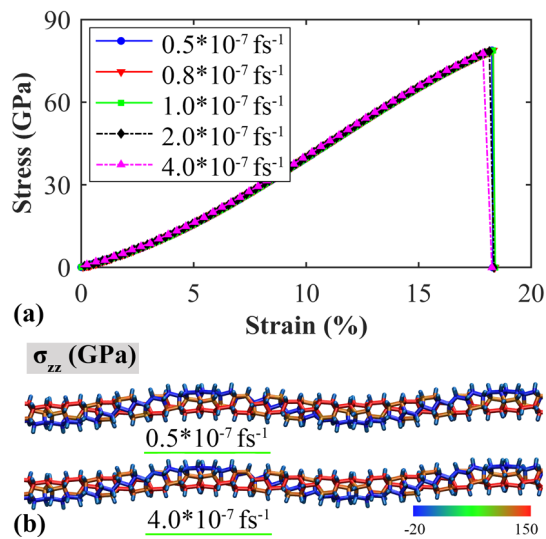

**Figure 3** (a) Comparisons of the stress-strain curves of NTH-III under different strain rates at 1 K. (b) The atomic stress distribution of NTH-III at the strain of 12% along the stretch direction. Here the strain rates are $0.5 \times 10^{-7}$ (upper) and $4.0 \times 10^{-7}$ fs$^{-1}$ (lower), respectively. Atoms are coloured according to the atomic stress along the length direction.

*3.2 Temperature impacts*

As mentioned above, the NTH is normally comprised of by pentagon, hexagon, heptagon and octagon carbon rings, which leads to a relatively large carbon bond length ranging from 1.51 to 1.67 Å (at 0 K) [26]. With these long and non-uniformly distributed carbon bonds, it is crucial to understand the thermal influence on the mechanical properties of the NTH. We have examined the tensile deformation of the three selected NTHs under temperatures ranging from 1 to 300 K.

As illustrated in Figure 4, the temperature is found to exert a significant impact on the mechanical performance of the NTH. All three NTHs have a smaller failure strength/strain at higher temperature. For the stiff-chiral NTH-II (Figure 4b), the yield strength at 300 K (~ 117 GPa) is about 18% smaller than that at 1 K. Similarly, the failure strain at 300 K (~ 14.7%) is over 45% smaller than that at 1 K. Such reduction is also seen in the case of the achiral NTH-I and soft-chiral NTH-III. From Figure 4c, the failure strength of NTH-III decreases from ~ 79 GPa to ~ 27 GPa when the temperature changes from 1 K to 300 K, corresponding to more than 60% reduction.

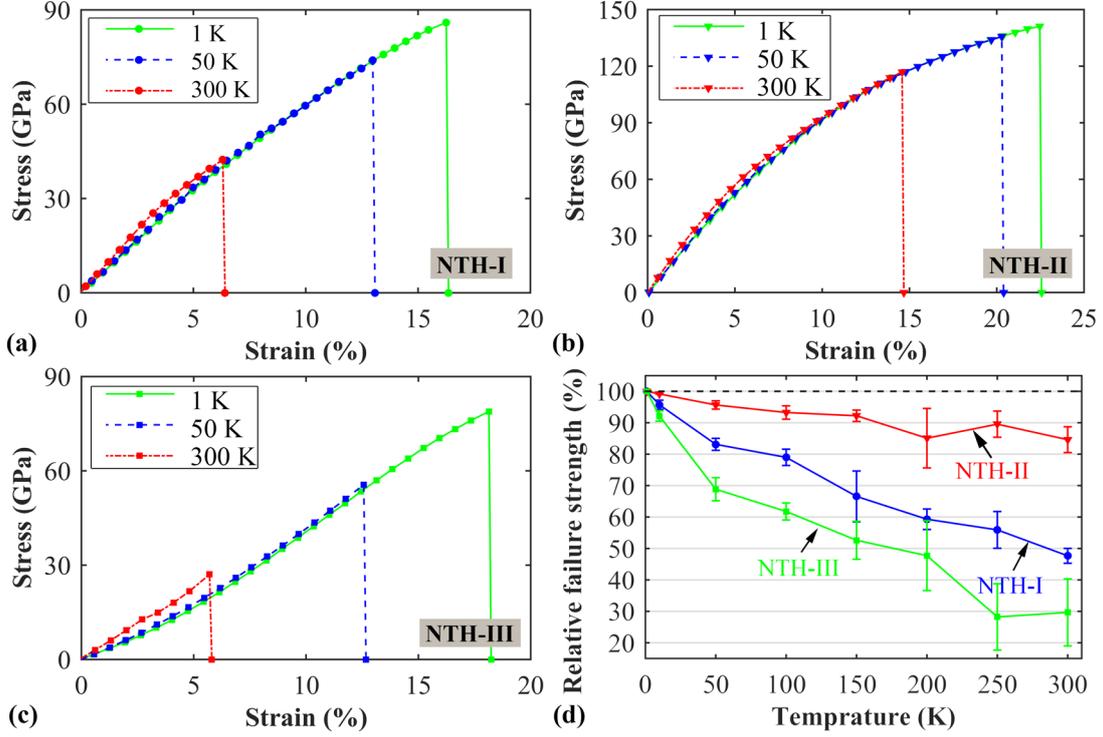

**Figure 4** Comparisons of the stress-strain curves at the temperature of 1, 50 and 300 K for: (a) NTH-I, (b) NTH-II, and (c) NTH-III; (d) The relative failure strength of the three NTHs as a function of temperature. Errorbar represents the relative standard deviation of the failure strength calculated from four different simulations with relaxation time ranging from 1 to 2.5 ns.

Figure 4d shows the relative failure strength ($\sigma_{fr}$) as a function of temperature. Here $\sigma_{fr} = \sigma_{fn}/\sigma_{f0}$, with $\sigma_{fn}$ and $\sigma_{f0}$ represent the failure strength at the temperature of *n* and 1 K, respectively. The failure strength is a mean value averaged over four different simulations with relaxation time ranging from 1 to 2.5 ns. Different relaxation time is adopted here to endow the NTH with slightly different initial status before tensile deformation. Our simulations have shown that these four simulations yield to similar stress-strain curves for each case. According to Figure 4d, although there are certain fluctuations, it is clearly seen that the failure strength of the NTH decreases with the increase of the temperature. Comparing with the other two NTHs, the stiff-chiral NTH-II has the least reduction (e.g., ~ 15% at 300 K) while the temperature increases from 1 K to 300 K. The most remarkable reduction is seen in the case of NTH-III (~ 70% at 300 K). Meanwhile, the relative failure strain of the NTH shows a similar trend (see Supporting Information). Unlike failure strength/strain, the effective Young's modulus appears to be independent of temperature (see Supporting Information). Within the examined temperature range, the Young's modulus is found to fluctuate around 0.69 ± 0.03, 1.17 ± 0.05, and 0.30 ± 0.03 TPa for NTH-I, NTH-II, and NTH-III, respectively.

To understand the significant temperature impacts on the mechanical properties of NTHs, we qualitatively compare the carbon bond length distribution of the NTH at different temperatures. To achieve this, the distribution of the carbon bond length is recorded every 10 ps for a total of 1 ns during the relaxation process (after the system reaches an equilibrium state). These time series values were then used to derive a time averaged distribution and cumulative density function (CDF) of the carbon bond length. Figure 5a shows the CDF of the carbon bond length of NTH-III at the temperature of 1, 50 and 300 K (see Supporting Information for the corresponding bond distribution and also the results for NTH-I and NTH-II). It is seen that the carbon bond length has a larger range at higher temperature, indicating both bond shortening and lengthening at increased temperatures. For NTH-III (Figure 5a), the carbon bond length is about 1.558 ± 0.023 Å at 1 K, 1.560 ± 0.026 Å at 50 K, and 1.564 ± 0.039 Å at 300 K. Along with this change, the percentages of shorter or longer carbon bonds in the NTH also increase. As highlighted by the coloured regions in Figure 5a, the percentage of carbon bonds longer than ~ 1.61 Å at 300 K is much larger than that at 50 K. It is expected that the presence of longer carbon bonds, with correspondingly lower C-C bond strength, will lead to bond failure at lower stress, and thus lead to smaller failure strength/strain. This is similar to the change of mechanical properties in single-walled carbon nanotubes which was explained by coefficient of thermal expansion (CTE) [40].

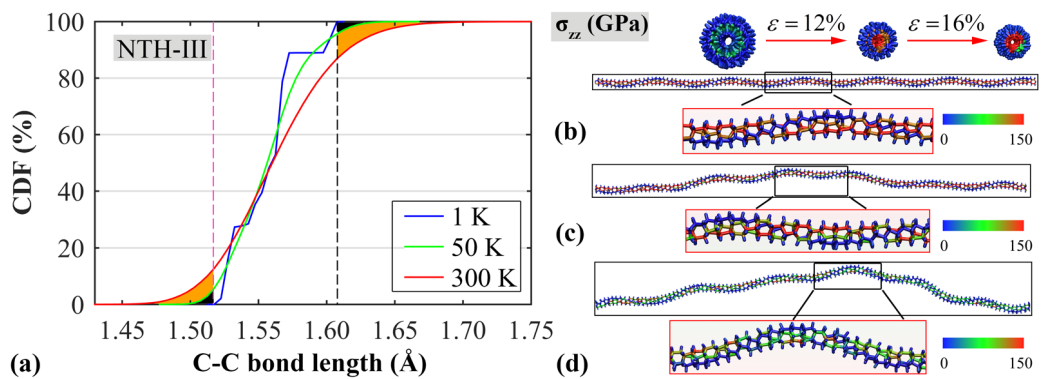

**Figure 5** (a) The cumulative density function (CDF) of the carbon bond length of NTH-III at the temperature of 1, 50 and 300 K. The left and right dash lines indicate the shortest and longest carbon bond length of NTH-III at 1 K, respectively; (b) The atomic stress distribution of NTH-III at the strain of 12% along the stretch direction under 1 K. Upper shows the shrinkage of the NTH's cross-section with increasing strain; The atomic stress distribution of NTH-III at the strain of: (c) 11% under 50 K and (d) 4% under 300 K, along the stretch direction. Atoms are coloured according to the atomic stress along the length direction.

Besides, the free lateral/bending vibration of the NTH at higher temperature is also a crucial factor that affects its mechanical performance. Take the NTH-III for an example, at low temperature (1 K), the sample can maintain its helical structure well during tensile deformation. As shown in Figure 5b, we can see a nice shrinkage of the NTH's cross-section at different strains (upper) and its axis is relatively straight. However, at higher temperature (50 K in Figure 5c), obvious offset of the NTH's axis is observed, which is resulted from the lateral vibration. With increasing temperature, such offset becomes more significant and starts to change the stress distribution of the NTH. As demonstrated in Figure 5d (at 300 K), the double-helix stress distribution pattern as observed at 1 K disappeared (Figure 2d and Figure 5b), instead, we only find some discrete stress concentration points. Similar results have also been observed from NTH-I and NTH-II. These findings have clearly shown that the thermal-induced free lateral vibration will greatly alter the mechanical properties of the NTH. Such obvious temperature impacts are also expected on other NTHs. Particularly, for the NTH with Stone-Wales transformation defects [19], higher temperature will alter the stress distribution and thus lead to different failure strength. Since the bond length is much longer around the Stone-Wales transformation defects than other portions, increasing defect density will increase the sensitivity of the properties to temperature.

*3.3 Bending rigidity*

Before concluding, we also assess the flexibility of the diamond NTH to further understand the mechanical characteristics of the three representative NTHs. Above results have shown that the DTH could undergo evident lateral vibration during tension, thus, a bending load is not appropriate for the bending stiffness calculation. Following previous work on similar one-dimensional carbon materials [19, 37], we impose different curvatures to bend the nanothread. Due to its helical morphology, the bending stiffness of the soft-chiral NTH-III is excluded from below discussion. A sample size around 8 nm was chose for both NTH-I and NTH-II. The curvature was introduced to the nanothread by bounding the NTH to an idealized surface (see Figure 6a) with the wall-atom interactions being described by a Lennard-Jones (LJ) 9/3 potential expressed as

$$E = \xi\left[\frac{2}{15}\left(\frac{\sigma}{r}\right)^9 - \left(\frac{\sigma}{r}\right)^3\right] \quad (2)$$

Here $\xi$ and $\sigma$ were chosen as 0.65 eV and 2 Å, respectively, following Roman *et al* [19]. The LJ 9/3 potential is derived by integrating over a 3D half-lattice LJ 12/6 particles, which effectively representing a semi-infinite LJ surface. The chosen potential depth $\xi$ has slightly increased the potential energy of the straight NTHs (upper Figure 6a, by ~ 0.001 and ~ 0.04 eV for NTH-I and NTH-II, respectively), suggesting ignorable local strain on the NTH due to the artificial surface. Curvatures range from 0.0083 to 0.033 Å$^{-1}$ were considered for the NTH. After energy minimization, the corresponding bending energy $E_b$ of NTH can be estimated by comparing its potential energy $E_r$ with that of a straight/unbent NTH adhered to the idealized wall $E_0$, i.e., $E_b = E_r - E_0$. Meanwhile, according to the continuum elastic theory, the elastic bending energy can be calculated from [41]

$$E_b = \frac{1}{2}D\rho^2 L \quad (3)$$

where $D$ is the bending stiffness, $\rho$ is the curvature, and $L$ is the sample length.

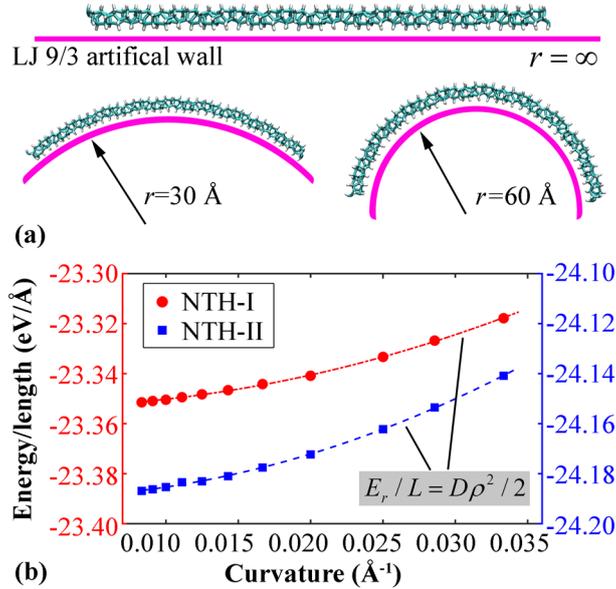

(a)

(b)

**Figure 6** (a) Schematic view of the NTH-II with different curvatures introduced by an artificial smooth surface, with the wall-atom interaction depicted by Lennard-Jones 9/3 potential; (b) Potential energy normalized by sample length $E_r / L$ as a function of the curvature for NTH-I and NTH-II.

Figure 6b shows the normalized potential energy per length ($E_r/L$) for NTH-I and NTH-II as a function of the curvature. The parabolic dependence of $E_r/L$ on the curvature $\rho$ is in line with the relationship described by Eq. 3. From linear regression, the bending stiffness $D$ is about 64 eV·Å (1480 kcal/mol·Å) and 88 eV·Å (2030 kcal/mol·Å) for NTH-I and NTH-II, respectively, which is higher than that estimated for the diamond nanothread with Stone-Wales transformation defects (~ 770 kcal/mol·Å) [19]. In comparison, a (5,5) CNT has a much higher rigidity on the order of 100 000 kcal/mol·Å [19], and cumulene carbyne chain and polyyne carbyne chain have a bending rigidity around 8.5 and 6.7 eV·Å [37], respectively. With the estimated bending rigidity, the persistence length, which is an important parameter for use of NTH as a structural connection, can be calculated. Following the concept in polymer physics [42], the persistence length $L_p$ can be estimated from $L_p = D/k_B T$, where $k_B$ is the Boltzmann constant and $T$ is the temperature. The persistence length for NTH-I and NTH-II is found to be about ~250 and ~340 nm at 300 K, respectively. These results indicate that a NTH (I or II) shorter than $L_p$ behaves like a flexible elastic rod, while for a NTH longer than $L_p$, its properties can only be described statistically. Compared to polymers ($L_p \approx 1$ nm) or double-stranded DNA ($L_p \approx 45$ - 50 nm) [43], NTH-I and NTH-II are relatively rigid. While, compared to CNTs ($L_p \approx 10$ - 100 μm), they are relatively flexible. Since the longest NTH synthesised in experiment is only 24 nm, thus, they behave more like a flexible elastic rod.

4. **Conclusions**

In summary, we have modelled the tensile behaviour of three representative diamond NTHs through MD simulations. It is found that the mechanical properties of NTH can vary significantly due to the morphology differences. For instance, the failure strength and Young's modulus for the studied stiff-chiral NTH are as high as 141 GPa and 1.09 TPa at 1 K, respectively. In comparison, the soft-chiral NTH shows much smaller failure strength of about 79 GPa and Young's modulus of 0.29 TPa. Such big difference is supposed as originated from the different stress distribution under tensile strain that is determined by its structure. It is found that the two NTHs that exhibit much lower failure strength/strain and Young's modulus exhibit evident stress concentration during tensile deformation. Further studies have shown that the temperature exerts a significant impact on the mechanical properties of the NTH. Specifically, the failure strength/strain decreases with the increase of temperature, and the effective Young's modulus appears irrelevant to temperature. The remarkable reduction of the failure strength/strain is regarded as resulted from the expansion of the structure and also the free lateral vibration that triggered at higher temperature. Additional calculations have shown that the NTH (-I and –II) has smaller bending rigidity than CNT but much higher than that of the carbyne chain, suggesting that it behaves like flexible elastic rods. This study has provided a fundamental understanding of the tensile behaviours of different NTHs and also elucidated the temperature impacts, which should shed lights on the design and also application of the NTH-based nanostructures. Particularly, the highly tailorable characteristics of the structure and mechanical properties of NTH are expected to show broad applications for the cross-linked systems, such as NTH-based yarn or fiber. In this respect, a comprehensive understanding of its torsional properties is a necessity, which will be the focus of our subsequent study.

**AUTHOR INFORMATION**

**Acknowledgement**

Supports from the ARC Discovery Project (DP130102120) and the High Performance Computer resources provided by the Queensland University of Technology are gratefully acknowledged.

**Supporting Information**

Supporting information is available for the tensile deformation of NTH-III under a constant velocity tensile loading, the temperature influence on the relative failure strain and Young's modulus of NTH, and the distribution function and cumulative density function of carbon bond length of NTH-I and NTH-II at different temperatures.

**References**

[1] Iijima S. Helical microtubules of graphitic carbon. Nature 1991; 354(6348): 56-58.
[2] Sun L, Gong J, Zhu D, Zhu Z, He S. Diamond nanorods from carbon nanotubes. Adv. Mater. 2004; 16(20): 1849-1853.
[3] Geim AK. Graphene: status and prospects. Science 2009; 324(5934): 1530-1534.
[4] Moser J, Eichler A, Güttinger J, Dykman MI, Bachtold A. Nanotube mechanical resonators with quality factors of up to 5 million. Nat Nano 2014; 9: 1007-1011.
[5] Chaste J, Eichler A, Moser J, Ceballos G, Rurali R, Bachtold A. A nanomechanical mass sensor with yoctogram resolution. Nat. Nanotechnol. 2012; 7(5): 301-304.
[6] Liu Z, Fang S, Moura F, Ding J, Jiang N, Di J; et al. Hierarchically buckled sheath-core fibers for superelastic electronics, sensors, and muscles. Science 2015; 349(6246): 400-404.
[7] Cheng H, Hu C, Zhao Y, Qu L. Graphene fiber: a new material platform for unique applications. NPG Asia Mater 2014; 6: e113.
[8] Ghosh T. Stretch, wrap, and relax to smartness. Science 2015; 349(6246): 382-383.
[9] Yang J, Ma M, Li L, Zhang Y, Huang W, Dong X. Graphene nanomesh: new versatile materials. Nanoscale 2014; 6(22): 13301-13313.
[10] Slater AG, Cooper AI. Function-led design of new porous materials. Science 2015; 348(6238): aaa8075.
[11] Shenderova O, Brenner D, Ruoff RS. Would diamond nanorods be stronger than fullerene nanotubes? Nano Lett. 2003; 3(6): 805-809.
[12] Yu Y, Wu L, Zhi J. Diamond Nanowires: Fabrication, Structure, Properties, and Applications. Angew. Chem. Int. Ed. 2014; 53(52): 14326-14351.
[13] Shang N, Papakonstantinou P, Wang P, Zakharov A, Palnitkar U, Lin IN; et al. Self-Assembled Growth, Microstructure, and Field-Emission High-Performance of Ultrathin Diamond Nanorods. ACS Nano 2009; 3(4): 1032-1038.
[14] Hsu C-H, Xu J. Diamond nanowire - a challenge from extremes. Nanoscale 2012; 4(17): 5293-5299.
[15] Coffinier Y, Szunerits S, Drobecq H, Melnyk O, Boukherroub R. Diamond nanowires for highly sensitive matrix-free mass spectrometry analysis of small molecules. Nanoscale 2012; 4(1): 231-238.
[16] Babinec TM, Hausmann BJ, Khan M, Zhang Y, Maze JR, Hemmer PR; et al. A diamond nanowire single-photon source. Nat. Nanotechnol. 2010; 5(3): 195-199.
[17] Yang N, Uetsuka H, Osawa E, Nebel CE. Vertically aligned diamond nanowires for DNA sensing. Angew. Chem. Int. Ed. 2008; 47(28): 5183-5185.
[18] Yang N, Uetsuka H, Nebel CE. Biofunctionalization of Vertically Aligned Diamond Nanowires. Adv. Funct. Mater. 2009; 19(6): 887-893.
[19] Roman RE, Kwan K, Cranford SW. Mechanical Properties and Defect Sensitivity of Diamond Nanothreads. Nano Lett. 2015; 15: 1585-1590.
[20] Fitzgibbons TC, Guthrie M, Xu E-s, Crespi VH, Davidowski SK, Cody GD; et al. Benzene-derived carbon nanothreads. Nat. Mater. 2015; 14(1): 43-47.
[21] Stojkovic D, Zhang P, Crespi VH. Smallest nanotube: Breaking the symmetry of sp 3 bonds in tubular geometries. Phys. Rev. Lett. 2001; 87(12): 125502.
[22] Zhang J, Zhu Z, Feng Y, Ishiwata H, Miyata Y, Kitaura R; et al. Evidence of diamond nanowires formed inside carbon nanotubes from diamantane dicarboxylic acid. Angew. Chem. Int. Ed. 2013; 52(13): 3717-3721.
[23] Chen B, Hoffmann R, Ashcroft NW, Badding J, Xu E, Crespi V. Linearly Polymerized Benzene Arrays As Intermediates, Tracing Pathways to Carbon Nanothreads. J. Am. Chem. Soc. 2015; 137(45): 14373-14386.
[24] Olbrich M, Mayer P, Trauner D. A step toward polytwistane: synthesis and characterization of C 2-symmetric tritwistane. Org. Biomol. Chem. 2014; 12(1): 108-112.
[25] Barua SR, Quanz H, Olbrich M, Schreiner PR, Trauner D, Allen WD. Polytwistane. Chemistry-A European Journal 2014; 20(6): 1638-1645.
[26] Xu E-s, Lammert PE, Crespi VH. Systematic enumeration of sp3 nanothreads. Nano Lett. 2015; 15: 5124-5130.
13